\documentclass[prb,amsmath,amssymb,floatfix,twocolumn,citeautoscript]{revtex4}
\usepackage{bm}
\usepackage{epsfig}

\begin{document}

\title{Berezinskii-Kosterlitz-Thouless Transition in Heavy Fermion Superlattices}

\author{Jian-Huang She$^1$, and Alexander V. Balatsky$^{1,2}$}

\affiliation{$^1$Theoretical Division, Los Alamos National Laboratory, Los Alamos, NM, 87545, USA.\\
$^2$Center for Integrated Nanotechnologies, Los Alamos National Laboratory, Los Alamos, NM, 87545, USA.}

\begin{abstract}
We propose an explanation of the superconducting transitions discovered in the heavy fermion superlattices by Mizukami et al. (Nature Physics 7, 849 (2011)) in terms of Berezinskii-Kosterlitz-Thouless transition. We observe that the effective mass mismatch between the heavy fermion superconductor and the normal metal regions provides an effective barrier that enables quasi 2D superconductivity in such systems. We show that the resistivity data, both with and without magnetic field, are consistent with BKT transition. Furthermore, we study the influence of a nearby magnetic quantum critical point on the vortex system, and find that the vortex core energy can be significantly reduced due to magnetic fluctuations. Further reduction of the gap with decreasing number of layers is understood as a result of pair breaking effect of Yb ions at the interface.

\end{abstract}

\date{\today \ [file: \jobname]}

 \maketitle

 Thin film growth technology recently has advanced to the point that artificial two-dimensional structures can be fabricated with atomic-layer precision. This has enabled the exploration of novel aspects of emergent phenomena in low dimensional systems with unprecedented control. Using the molecular beam epitaxy (MBE) technique, Mizukami et al. have grown CeCoIn5/YbCoIn5 superlattices, where superconductivity was found to occur in the two-dimensional Kondo lattice  [\onlinecite{Mizukami11}]. The combination of f-electron physics, low dimensionality and interface effects provides a rare opportunity to study new states in strongly correlated electron systems, e.g. unconventional superconductivity, dimensionally-tuned quantum criticality [\onlinecite{Shishido10}], interplay of magnetism and superconductivity, Fulde-Ferrell-Larkin-Ovchinnikov phases, and to induce symmetry breaking not available in the bulk like locally broken inversion symmetry [\onlinecite{Sigrist12}].

Here, we investigate the mechanism for the onset of superconductivity in such heavy fermion superlattices. We propose an explanation of the experimental results of [\onlinecite{Mizukami11}] within the framework of Berezinskii-Kosterlitz-Thouless (BKT) transition, and further study the interplay of Kondo lattice physics and BKT mechanism. While well established for superfluid films, BKT transition is less convincing for superconductors (See [\onlinecite{Minnhagen87}] and references therein). Though implications have been found in numerous  thin superconducting films [\onlinecite{Minnhagen87, Firoy88, Davis90, Matsuda93, Crane07}], highly anisotropic cuprates [\onlinecite{Wen98, Corson99, Ong05}], oxide interfaces [\onlinecite{Reyren07, Caviglia08, Schneider09}], the results have remained inconclusive (see e.g. [\onlinecite{Kogan07, Benfatto09}]). It is therefore desirable to have a well-controlled, readily-tunable system to investigate the BKT physics. The epitaxially grown heavy fermion superlattices may serve such a role.

{\it Quasi 2-dimensional superconductivity:} First, we discuss why BKT theory is applicable to heavy fermion superlattices. In the CeCoIn$_5$/YbCoIn$_5$ superlattice, one has a layered structure of alternating heavy fermion superconductor (CeCoIn$_5$) and conventional metal (YbCoIn$_5$), typically 3.5 nm thick. Proximity effect is expected to happen in such normal metal/superconductor (N/S) junctions. For conventional superconductors, the thickness of the leakage region is on the order of the thermal length $\hbar v_{N}/2\pi k_B T$, where $v_{N}$ is the Fermi velocity in the N region (see e.g. [\onlinecite{deGennes69}] ). At low temperatures, this thickness is typically of order $100 nm$, which is much larger than the separation of CeCoIn$_5$ layers. One may thus expect a strong coupling between the superconducting CeCoIn$_5$ layers and the system would behave as three dimensional superconductor. However, as we will argue below, the large mismatch of Fermi velocities across the interface changes the story completely and enables quasi 2D superconductivity in CeCoIn$_5$ thin layers.

In normal metal/heavy fermion superconductor proximity effect studies, it was realized that the large mismatch of effective mass at the interface leads to huge suppression of transmission of electron probability currents [\onlinecite{Fenton85}]. The ratio $r_T$ of the transmitted probability current and the incident current is determined by the ratio of the effective masses,  $r_T\simeq 4m_l/m_h$, for $m_h\gg m_l$ [\onlinecite{Fenton85}]. The effective mass of CeCoIn$_5$ is of order $100 m_e$. For the more conventional metal YbCoIn$_5$, we take its effect mass to be of order $m_e$. The transmission is thus on the order of one percent. 

This result is intimately related to that of Blonder, Tinkham and Klapwijk [\onlinecite{BTK82,Blonder83}], where it was shown that the mismatch of Fermi velocities between the N and S regions increases the barrier height between the two, with the effective barrier parameter $Z$ modified to $Z=(Z_0^2+(1-r)^2/4r)^{1/2}$ where $r=v_S/v_N$ is the ratio of two Fermi velocities. This gives essentially the same result as Ref. [\onlinecite{Fenton85}]. This suppression factor significantly degrades the proximity coupling to the point where 4 nm normal layer renders heavy fermion films essentially uncoupled. A direct consequence of the reduced proximity effect is an enhanced c axis resistivity, which can be measured directly in experiment. 

More extensive numerical studies of proximity effect in N/S junctions have been carried out recently [\onlinecite{Valls10}], where it was shown that proximity effect is substantially suppressed with moderate mismatch of Fermi energies. Another source of suppression of the proximity effect is the pair breaking effects of Yb ions at the interface (see supplementary material). It is also expected that a weak magnetic field can destroy the proximity-induced superconductivity in YbCoIn$_5$ layers [\onlinecite{Mizukami11,Sefafin10}]. 

Suppression of the proximity effect in the CeCoIn$_5$/YbCoIn$_5$ superlattice and the fact that the thickness of the CeCoIn$_5$ layers is on the order of the perpendicular coherence length $\xi_{\perp} \sim 20 {\rm \AA}$  [\onlinecite{Mizukami11}], lead to the conclusion that superconductivity in such systems is essentially two dimensional, and one expects BKT physics to be relevant in such systems.

\begin{figure}
\begin{centering}
\includegraphics[width=0.6\linewidth]{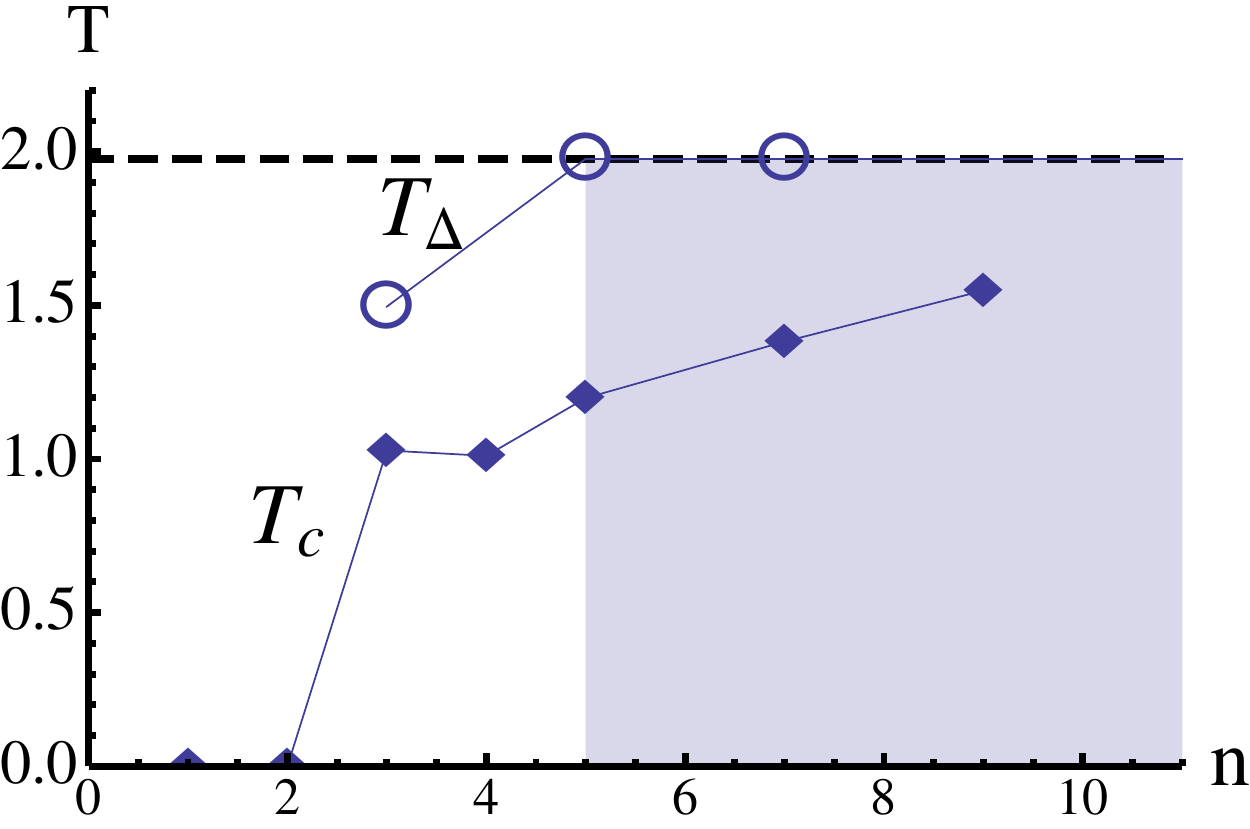}
\end{centering}
\caption{Gap and $T_c$ as function of number of CeCoIn$_5$ layers $n$ (data from Mizukami et al. [\onlinecite{Mizukami11}]). For $n\geq 5$ (shaded region), gap retains the bulk value, while $T_c$ decreases with decreasing number of layers.}
\label{gTc}
\end{figure}

{\it BKT transition:} The basic experimental fact of Mizukami et.al [\onlinecite{Mizukami11}] is that when the number of CeCoIn$_5$ layers $n\geq 5$, the upper critical field $H_{c2}$, both parallel and perpendicular to the ab-plane, retains the bulk value, while the transition temperature $T_c$ decreases with decreasing $n$ (see Fig.\ref{gTc}). $H_{c2}$ in such systems is Pauli-limited in both parallel and perpendicular directions [\onlinecite{Mizukami11, Bianchi08}] and is thus a direct measure of the superconducting gap, with $H_{c2}^{\rm Pauli}\simeq \sqrt{2}\Delta/g\mu_B$,  where $g$ is the gyromagnetic factor and $\mu_B$ is the Bohr magneton. This means that gap retains the bulk value for $n\geq 5$. The behavior of gap and $T_c$ for different number of CeCoIn$_5$ layers is shown in Fig.~\ref{gTc}. Our proposal is that such behavior is due to the effect of phase fluctuations, which for the quasi-two-dimensional superconductors considered here is controlled by the Berezinskii-Kosterlitz-Thouless physics [\onlinecite{Berezinskii70, Kosterlitz73}].

For two dimensional systems with continuous Abelian symmetry, despite the lack of broken symmetry due to strong fluctuations, there exists a finite temperature phase transition mediated by topological defects, e.g. vortices for superconductors [\onlinecite{Berezinskii70, Kosterlitz73}]. Below the transition temperature $T_{\rm BKT}$, vortices and antivortices are bound into pairs, and the resistance vanishes. Above $T_{\rm BKT}$, vortex-antivortex pairs unbind, and the proliferation of free vortices destroys superconductivity. For such systems, one thus has $T_c=T_{\rm BKT}$.

For layered superconductors, one also needs to include interlayer couplings. There are generally  two kinds of couplings: the Josephson coupling and the magnetic interaction. Since the separation of the different CeCoIn$_5$ layers is larger than the perpendicular coherence length, the interlayer Josephson coupling is weak, and can be ignored. The long range magnetic interaction couples vortices in different planes, and aligns vortices of the same sign into stacks. Since the interlayer coupling is still logarithmic as in two dimensional superconductors, the phase transition is expected to remain in the same universality class as BKT transition [\onlinecite{Korshunov90}]. This has been confirmed by detailed renormalization group studies [\onlinecite{Horovitz92, Scheidl92, Horovitz93, Sondhi09}] (see also [\onlinecite{Timm95}]). It has also been shown in Ref. [\onlinecite{Sondhi09}] that $T_c$ is only slightly modified. \footnote{With $s\ll \lambda_{\parallel}$, the transition temperature now reads $T_c=(\pi/2)\rho_s(1-\frac{s}{2\lambda_{\parallel}})$, where $s$ is the layer spacing,  $\lambda_{\parallel}$ is the in-plane penetration depth, and $\rho_s=\Phi_0^2s/(16\pi^3\lambda_{\parallel}^2)$ is the in-plane superfluid stiffness, which can be measured directly. In the experiment of Mizukami et.al [\onlinecite{Mizukami11}], $s\sim 3.7nm, d\sim 5nm$. Taking $\lambda\sim \lambda_b\sim 358nm$, we have $\lambda_{\parallel}\sim 308$ and $s/2\lambda_{\parallel}\sim 0.006$.} While such small modification may be detected by future high precision measurements, as first approximation we will ignore it in the following and concentrate on the single-layer problem.

\begin{figure}
\begin{centering}
\includegraphics[width=0.48\linewidth]{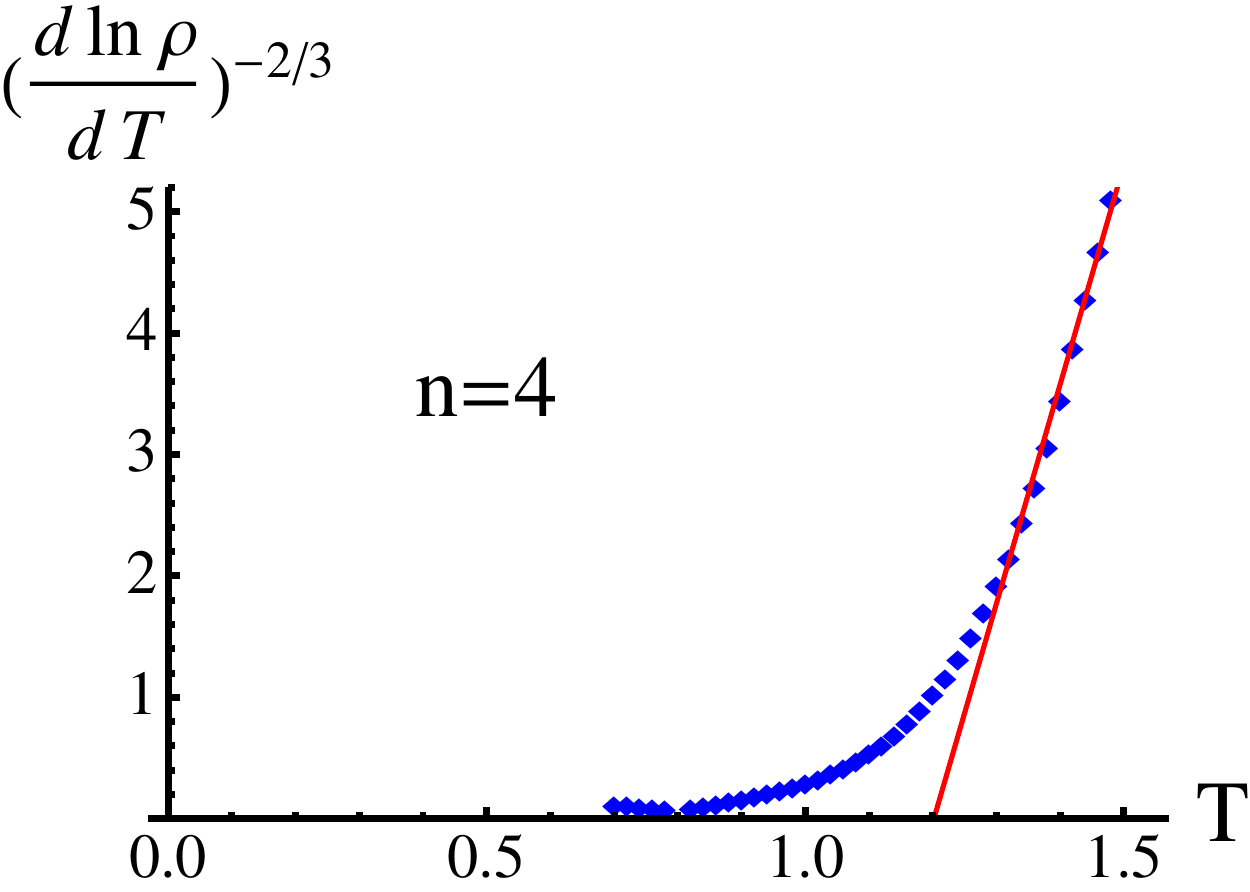}
\includegraphics[width=0.48\linewidth]{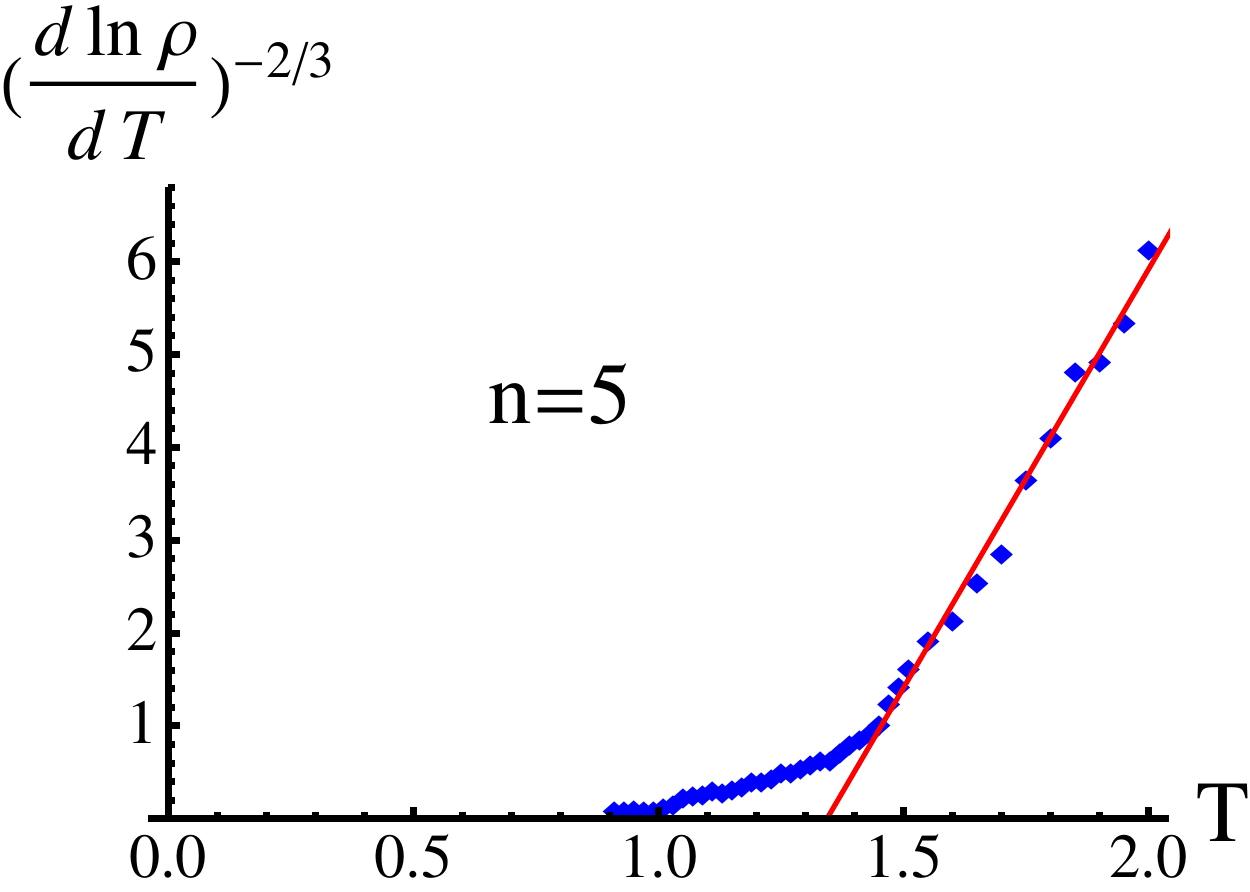}
\includegraphics[width=0.48\linewidth]{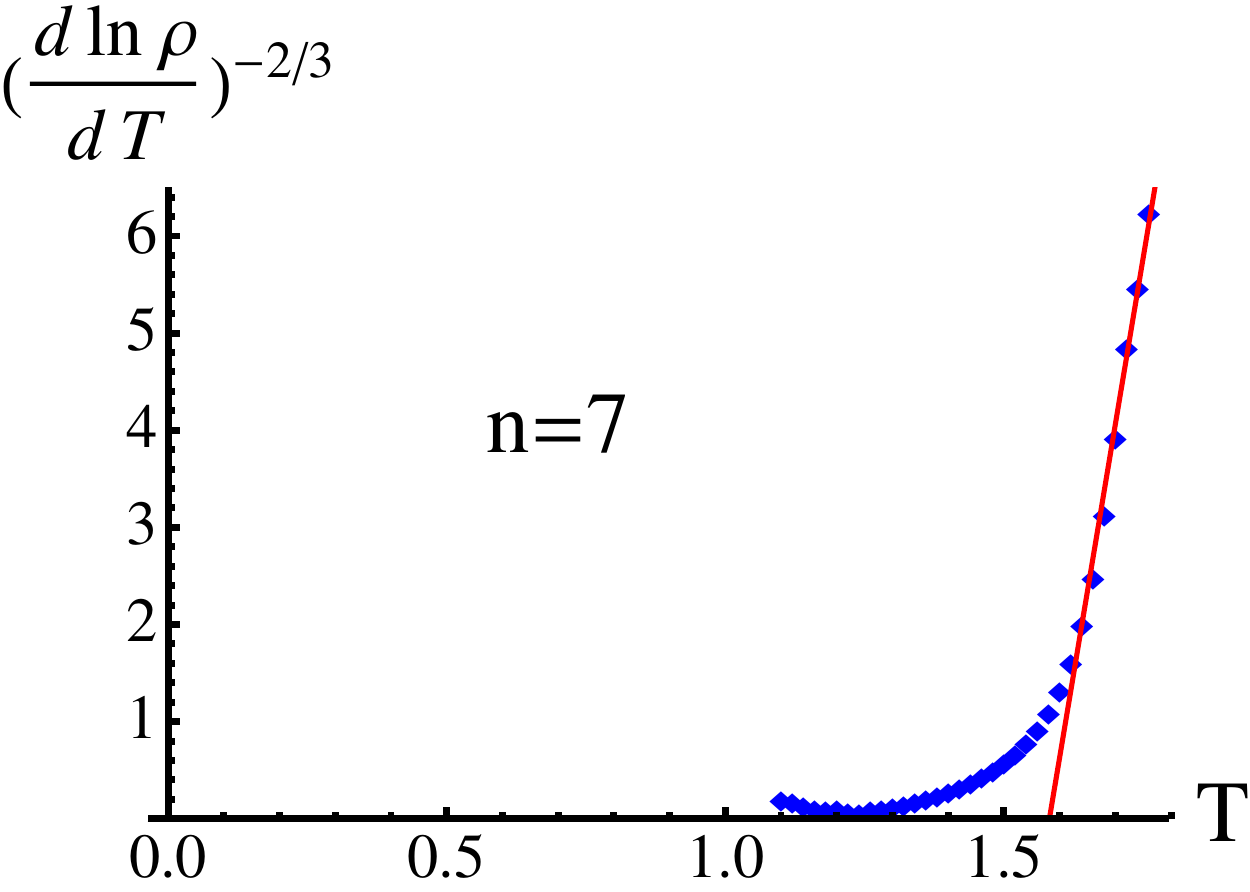}
\includegraphics[width=0.48\linewidth]{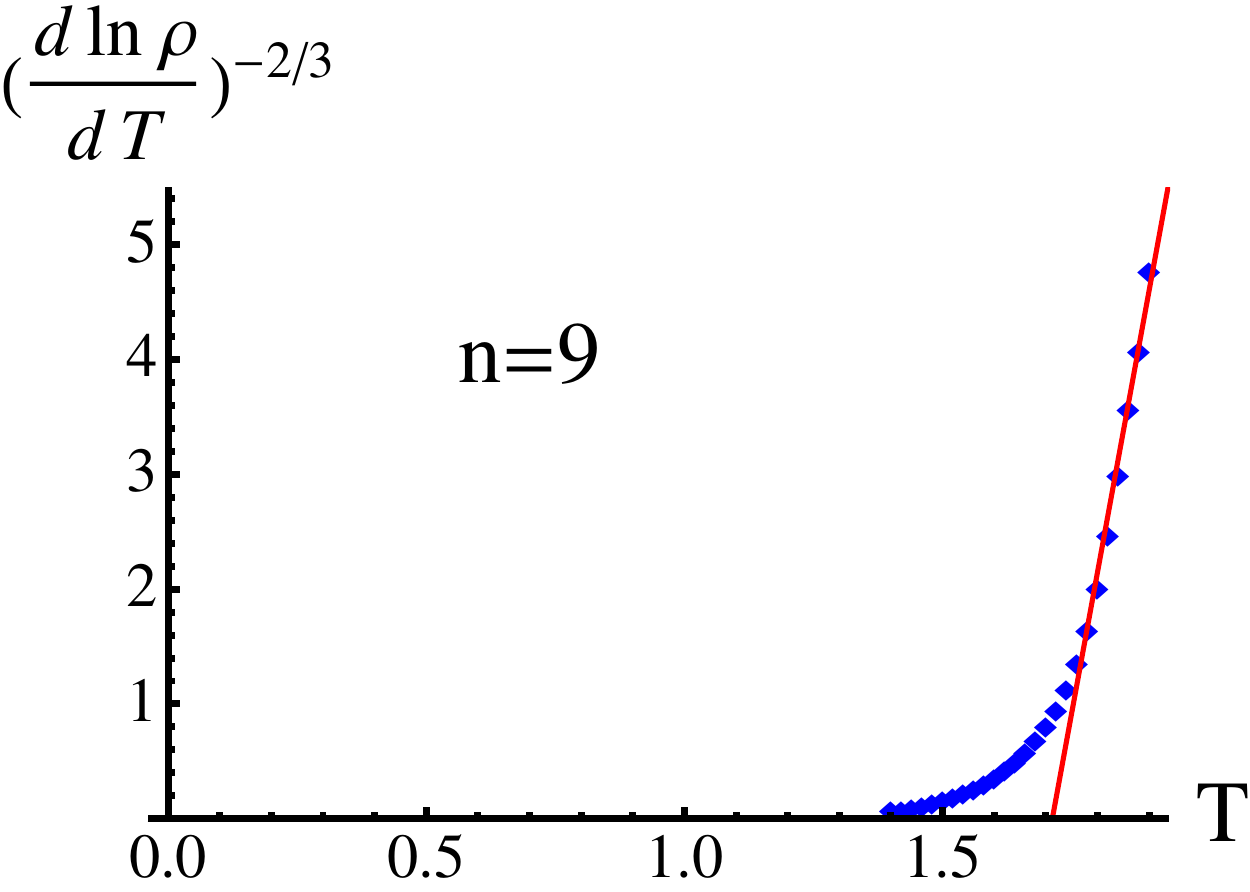}
\end{centering}
\caption{Resistivity as function of temperature for $n=4, 5, 7, 9$ (data from Mizukami et al. [\onlinecite{Mizukami11}]). The BKT transition temperature is determined from the intersection with the T-axis to be $T_{\rm BKT}=1.202, 1.344,
1.582, 1.712 K$ respectively. }
\label{res}
\end{figure}

\begin{figure}
\begin{centering}
\includegraphics[width=0.48\linewidth]{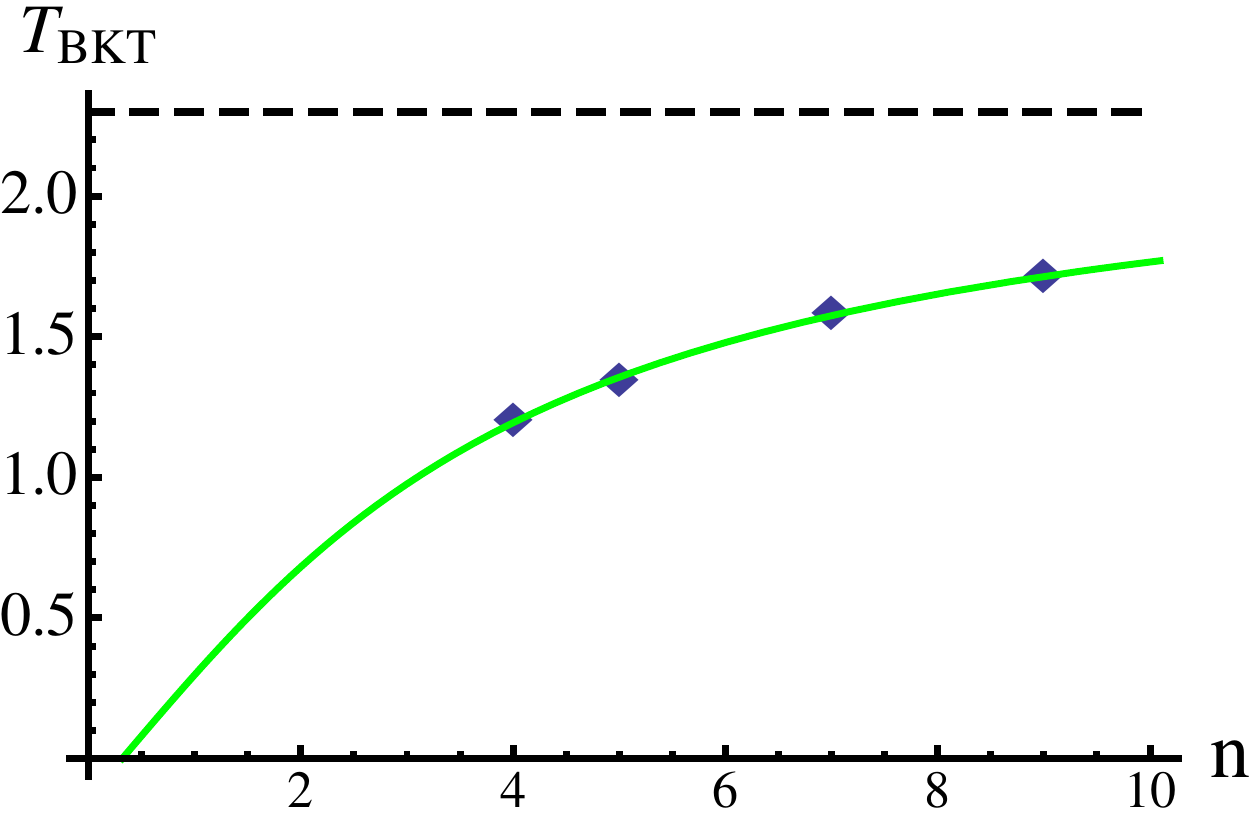}
\includegraphics[width=0.48\linewidth]{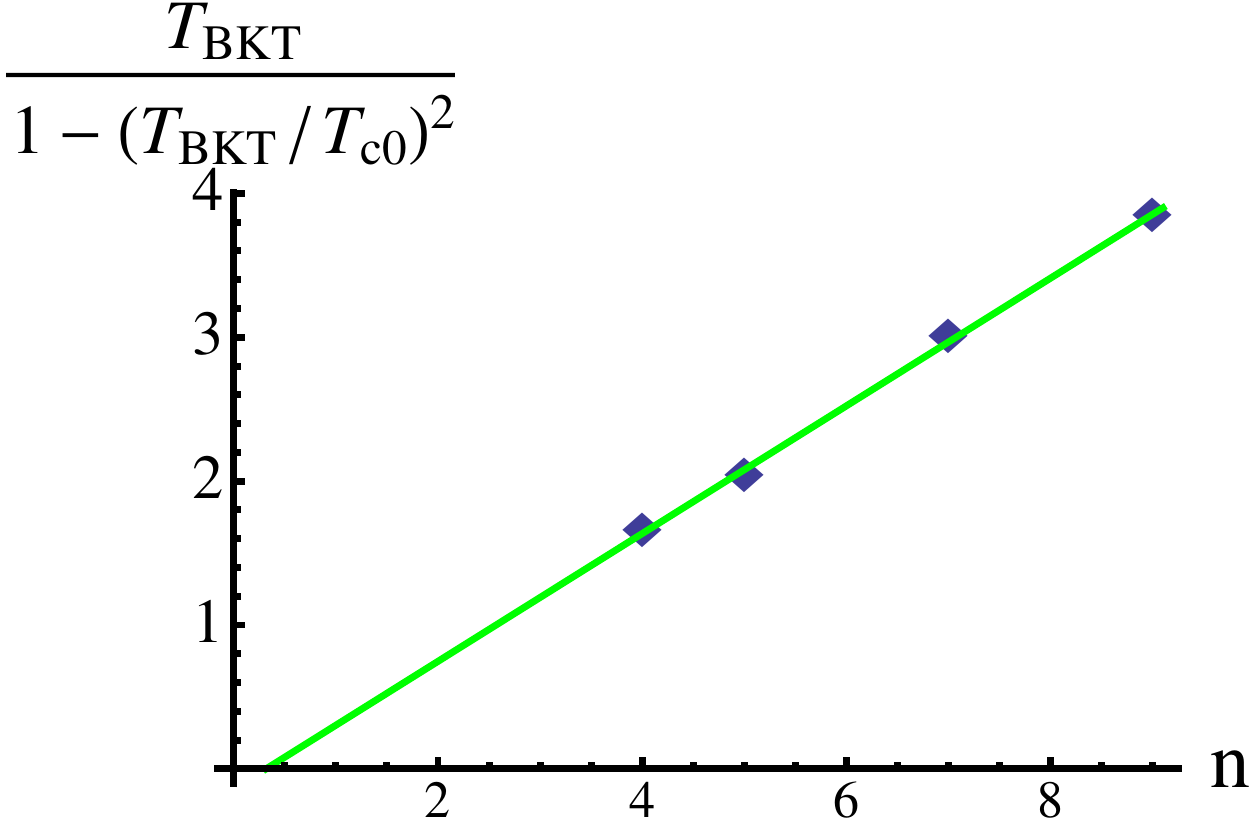}
\end{centering}
\caption{The BKT transition temperature $T_{\rm BKT}$ as function of the number of CeCoIn$_5$ layers. The dashed line is $T_{c0}=2.3K$. The solid line is a fit to the theoretical result, with $\frac{T_{\rm BKT}}{1-(T_{\rm BKT}/T_{c0} )^2}=0.444(n-0.317)$.}
\label{nTKT}
\end{figure}

In the following, we are going to check whether the experimental findings of Mizukami et al.  [\onlinecite{Mizukami11}] are consistent with BKT transition. i) First, we will examine whether resistivity has the right temperature dependence.  ii) Then we extract from the resistivity data the transition temperature $T_{\rm BKT}$. iii) Finally, we will check whether $T_{\rm BKT}$ has the right dependence on the number of layers. We find that the observations in [\onlinecite{Mizukami11}] are consistent with BKT transition.

Near $T_{\rm BKT}$, resistivity behaves as $\rho(T)=\rho_0 e^{ -b(T-T_{\rm BKT})^{-1/2}}$ [\onlinecite{Halperin79}], which gives $\left(d\ln\rho(T)/dT\right)^{-2/3}=\left(2/b\right)^{2/3} (T-T_{\rm BKT})$. We plot in Fig.~\ref{res} the temperature dependence of $(d\ln\rho(T)/dT)^{-2/3}$ for the four different cases with number of CeCoIn$_5$ layers $n=4, 5, 7, 9$, where one can see that $(d\ln\rho(T)/dT)^{-2/3}$ is indeed linear in $T$, and $T_{\rm BKT}$ can be extracted from the intersection points. We also notice that resistivity does not fall to zero at $T_{\rm BKT}$. It retains a small nonzero value in a temperature region below $T_{\rm BKT}$. This is generically observed for a BKT transition, and is attributed to the temperature difference between the formation of single vortices and the subsequent vortex condensation (see e.g. [\onlinecite{Bozovic11}] and references therein). Inhomogeneity and finite size effects also broaden the BKT transition, giving rise to the resistivity tail below $T_{\rm BKT}$ [\onlinecite{Benfatto09}].

Now, we proceed to study the thickness dependence of the BKT transition temperature.
$T_{\rm BKT}$ can be written as [\onlinecite{Kosterlitz73, Nelson77, Halperin79, Beasley79}]
\begin{equation}
k_BT_{\rm BKT}=\frac{\pi\hbar^2n_s^{2D}(T_{\rm BKT})}{8m\epsilon_c},
\end{equation}
with the dielectric constant $\epsilon_c\equiv n_s^{2D}/n_s^R$, where $n_s^R$ is the renormalized carrier density. The unrenormalized 2d carrier density $n_s^{2D}=n_s^{3D}d$ is determined by the 3d carrier density $n_s^{3D}(T)=n_s^{3D}(0)\lambda_b^2(0)/\lambda_b^2(T)$,
and the film thickness $d$. The bulk penetration depth $\lambda_b(T)$ has a temperature dependence of the form $\lambda_b(T)=\lambda_b(0)\left[ 1-\left( T/T_{c0} \right)^{\alpha}\right]^{-1/2}$,
with bulk mean field transition temperature $T_{c0}$. In the usual two-fluid picture, the exponent $\alpha=4$. For cuprates and CeCoIn$_5$, it has been found that $\alpha=2$ [\onlinecite{Bonn93, Kogan09}].
Thus we have
\begin{equation}
\frac{T_{\rm BKT}}{1-(T_{\rm BKT}/T_{c0} )^{2}}=\frac{\pi\hbar^2n_s^{3D}(0)}{8k_Bm\epsilon_c}d.
\label{TBKT1}
\end{equation}
Noting that $d=nx-d_0=(n-n_0)x$, with $n$ the number of CeCoIn$_5$ layers, $x$ the thickness of each layer and $d_0$ the thickness of the dead layers on top and bottom, the above result can be written as
\begin{equation}
\frac{T_{\rm BKT}{\rm [K]}}{1-(T_{\rm BKT}/T_{c0} )^{2}}=\frac{0.98 {\rm [cm]}\cdot x}{\lambda_b^2(0)}\frac{1}{\epsilon_c}(n-n_0).
\label{TBKT}
\end{equation}

We plot in Fig.~\ref{nTKT}  $T_{\rm BKT}$ as function of the number of CeCoIn$_5$ layers. The experimental results are in good agreement with the theoretical prediction determined from Eq.~\ref{TBKT}.
Taking $\lambda_b(0)=358 {\rm nm}$ [\onlinecite{Kogan09}], $x=\xi_c/4=2.1{\rm nm}/4$, we get the fitting parameter $\epsilon_c\simeq 90$.
With $\lambda^{-2}=\lambda_b^{-2}/\epsilon_c$, our prediction is that the penetration depth of the superlattice is enhanced by about one order of magnitude from the bulk value. Furthermore, another important prediction from BKT transition that can be checked is that the penetration depth of the superlattice $\lambda$ satisfies the universal relation [\onlinecite{Nelson77}]
\begin{equation}
k_BT_{\rm BKT} =\frac{\Phi_0^2}{32\pi^2}\frac{d}{\lambda^2},
\end{equation}
right below the transition temperature, where $\Phi_0=hc/2e$ is the flux quantum.

{\it Antiferromagnetic vortex core:}  We extract from the experiment [\onlinecite{Mizukami11}] a large dielectric constant $\epsilon_c$, which indicates a large fugacity, or a small vortex core energy  [\onlinecite{Kosterlitz73, Nelson77}] (see supplementary material for a more detailed analysis). Here, we try to understand where such a large renormalization may come from. We find that at the vortex core, where the superconducting gap is suppressed, magnetic ordering can occur locally (see e.g. [\onlinecite{Arovas97}]), which reduces the vortex core energy.

A salient feature of the heavy-fermion superconductor CeCoIn$_5$ is the proximity to an antiferromagnetic quantum critical point (QCP). Therefore, one may expect that fluctuating magnetic order may influence the vortex dynamics in the heavy fermion superlattices. Suppression of the superconductivity in the core can induce the antiferromagnetic state in the cores as opposed to a simple metal in conventional superconductors. To model this effect, we consider magnetic moment that couples to the vortex via a Zeeman term $g\mu_B H_v^zS^z$, where $H_v^z$ is the magnetic field generated by vortices.
 $H_v^z$ is a superpostion of the magnetic fields generated by vortices at different locations,  $H_v^z(\mathbf r)=\sum_i n_iH_0({\mathbf r}-{\mathbf R}_i)$, with $n_i$ the vorticity. $H_0({\mathbf r})$ can be obtained from its Fourier transform  $H_0(\mathbf k)=\Phi_0/(1+\lambda^2k^2)$, with result $H_0({\mathbf r}) \sim (\Phi_0/\lambda^2)K_0( r/\lambda)$,
where $K_0$ is the modified Bessel function of the second kind. For $r\ll \lambda$, $K_0\left( r/\lambda\right)\sim \ln r$.

Zeeman coupling induces a precession of the magnetic moment perpendicular to the magnetic field, which can be captured by modifying the kinetic energy density to $(\partial_{\tau}{\bm \phi}+ig\mu_B {\bm H}\times{\bm \phi} )^2$, where $\bm\phi$ is the sublattice magnetization density [\onlinecite{Affleck90, Affleck91, Rosch05}].
For ${\bm H}$ in the $z$-direction, one can define $\Phi=(\phi_x+i\phi_y)/\sqrt{2}$. Consider the static limit, its free energy density reads
\begin{equation}
{\cal F}_{\Phi}=|\nabla \Phi|^2+ (\alpha-g^2\mu_B^2H^2(r))|\Phi|^2+\gamma |\Phi|^4.
\end{equation}
Near the vortex core, $H\sim \ln |{\mathbf r}-{\mathbf r_i}|$ can be very large. Close to the QCP, $\alpha$ is small. When ${\tilde \alpha}\equiv \alpha-g^2\mu_B^2H^2<0$, the vortex core becomes antiferromagnetic, and  qualitatively $|\Phi|^2=-{\tilde \alpha}/2\gamma$ and the potential energy $V_{\Phi}=-{\tilde \alpha}^2/4\gamma<0$. Thus the vortex core energy is significantly reduced due to magnetic fluctuations.

\begin{figure}
\begin{centering}
\includegraphics[width=0.48\linewidth]{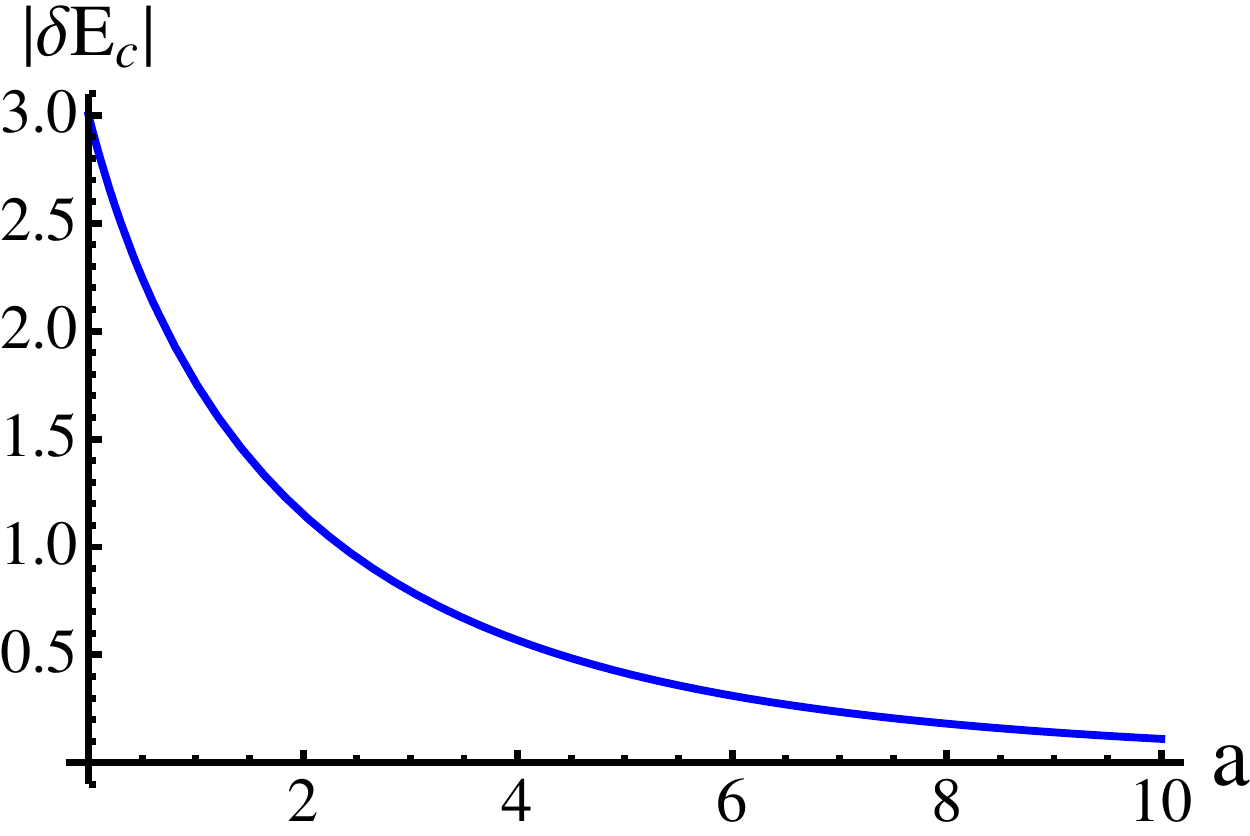}
\includegraphics[width=0.4\linewidth]{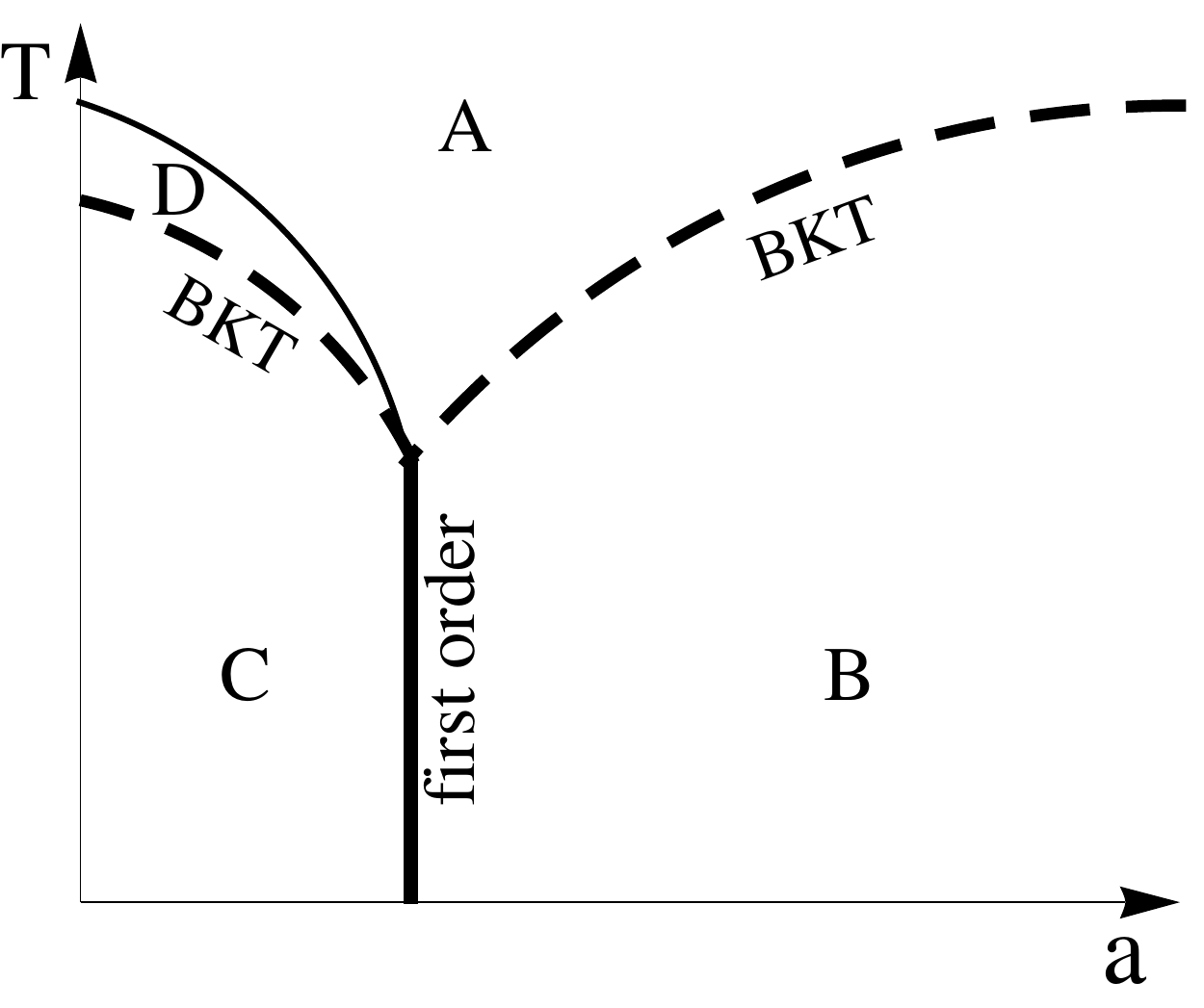}
\end{centering}
\caption{Change of vortex core energy as function of distance to the QCP and the resulting $a-T$ phase diagram as deduced from simulation [\onlinecite{Lee92}] and theoretical results [\onlinecite{Gabay93, Zhang93}] (see also [\onlinecite{Orkoulas96, Lidmar97}]). Phase A is gas of free vortices, B a gas of bound vortex-antivortex pairs, C a crystal of vortices and antivortices, D a hexatic phase of vortices and antivortices.}
\label{Ecore}
\end{figure}

More precisely, we consider the equation of motion
\begin{equation}
(-\nabla^2 + \alpha-g^2\mu_B^2H_0^2({\mathbf r})+2\gamma |\Phi({\mathbf r})|^2)\Phi({\mathbf r})=0,
\label{EoM}
\end{equation}
where a vortex of unit vorticity is placed at ${\mathbf r}=0$. Far away from the vortex core, i.e. $r\gtrsim\lambda$, $H_0$ decays exponentially, and $\Phi=0$ is the lowest energy solution. Near the vortex core, we can ignore $\alpha$ and $\Phi(r)\sim\ln (r/\lambda)$ is the lowest energy solution. The change of vortex core energy is $\delta E_c=\int d^2{\mathbf r}{\cal F}[\Phi({\mathbf r})]\sim -g^4\mu_B^4\Phi_0^4/\gamma\lambda^6\equiv-V_0<0$. For $\gamma$ small, core energy lowering effect can be very large.

We also notice that the vortex core energy depends on $\alpha$, the distance to the QCP. With the dimensionless quantity $a\equiv \alpha\lambda^4/g^2\mu_B^2\Phi_0^2$, the change of vortex core energy is $\delta E_c\sim -V_0\int_0^{r^*/\lambda} xdx(\ln^2x-a)^2$, where $r^*=\lambda e^{-\sqrt{a}}$ is the radius where magnetic condensate vanishes. And we have $\delta E_c\sim -V_0e^{-2\sqrt{a}}(3+6\sqrt{a}+4a)$ (see Fig.~\ref{Ecore}). One can thus tune the vortex fugacity by changing the distance to the QCP. It would be interesting to see whether phase diagrams as shown in Fig.~\ref{Ecore} can be observed experimentally.

{\it Effect of the magnetic field:} In the presence of a perpendicular magnetic field ($H\perp {\rm ab}$), there will be an imbalance of vortices parallel to the magnetic field and those anti-parallel, with $|n_+-n_-|>0$ [\onlinecite{Doniach79}]. The unbounded vortices will give rise to finite resistance. When the magnetic field is applied parallel to the $ab$-plane, there will be no such effects. This explains the enhanced resistivity when applying perpendicular magnetic field (Fig. 2c in [\onlinecite{Mizukami11}]). One can also see that a small parallel field will not change $T_{\rm BKT}$, i.e. $\partial T/\partial H_{c2\parallel}=0$ near $T_{\rm BKT}$, while a small perpendicular field will reduce $T_{\rm BKT}$, i.e. $\partial T/\partial H_{c2\perp}<0$ near $T_{\rm BKT}$, as observed in Fig. 4a of [\onlinecite{Mizukami11}]. Near $T_{\rm BKT}$, where both $H_{c2\parallel}$ and $H_{c2\perp}$ approach zero, the ratio $H_{c2\parallel}/H_{c2\perp}=(\partial T/\partial H_{c2\perp})/(\partial T/\partial H_{c2\parallel})$ thus diverges, as seen in Fig. 3b of [\onlinecite{Mizukami11}].

{\it Conclusions:} In conclusion, we have proposed that superconducting transition in the heavy fermion superlattice of Mizukami et al.[\onlinecite{Mizukami11}] is controlled by BKT transition of vortex-antivortex (un)binding. We have also shown that magnetic fluctuations modify the conventional BKT discussion since they reduce the vortex core energy, and thus quantum criticality may strongly influence the phase diagram of the vortex system. We made suggestions  to further test our proposal: The most clear signature of the BKT transition is a jump in the superfluid density at the transition [\onlinecite{Nelson77}], which can be detected by measuring the penetration depth. CeCoIn$_5$ sandwiched with insulating layers may make an even better two dimensional superconductor. In the opposite limit of a very thin normal YbCoIn$_5$ layer, we expect the crossover to conventional 3D superconducting transition that also would be interesting to test. In a dense vortex matter, vortex-antivortex pairs may crystallize, and subsequent melting may lead to intermediate hexatic phase[\onlinecite{Gabay93, Zhang93}]. It would be interesting to look for such phases in systems close to a magnetic QCP, where vortex core energy can be substantially reduced.

{\it Note added:} While this work was under review, we received a preprint by Fellows et al. [\onlinecite{Fellows12}], where they study a related problem of BKT transition in the presence of competing orders, focusing on the behavior near the high symmetry point.

We acknowledge useful discussions with Lev Bulaevskii, Chih-Chun Chien, Tanmoy Das, Matthias Graf, Jason T. Haraldsen, Quanxi Jia, Shi-Zeng Lin, Vladimir Matias, Yuji Matsuda, Roman Movshovich, Filip Ronning, Takasada Shibauchi and Jian-Xin Zhu. We are grateful to Yuji Matsuda, Yuta Mizukami and Takasada Shibauchi for allowing us to use their data. This work was supported, in part, by UCOP-TR01, by  the Center for Integrated Nanotechnologies, a U.S. Department of Energy, Office of Basic Energy Sciences user facility and in part by LDRD.  Los Alamos National Laboratory, an affirmative action equal opportunity employer, is operated by Los Alamos National Security, LLC, for the National Nuclear Security Administration of the U.S. Department of Energy under contract DE-AC52-06NA25396.

\section*{Supplementary Material}

\subsection{The dielectric constant} 

 Here we elaborate on the understanding of the dielectric constant $\epsilon_c$. In BKT theory, the vortex system is descibed by the Hamiltonian
\begin{eqnarray}
\frac{{\cal H}_v}{k_BT}=&-&\pi K\int d^2{\mathbf r}  \int d^2{\mathbf r'}n(\mathbf r)n(\mathbf r')\log\frac{|{\mathbf r}-{\mathbf r}'|}{R_0} \nonumber\\
&-&\log y \int d^2{\mathbf r} n^2(\mathbf r),
\end{eqnarray}
where the stiffness $K=n_s\hbar^2/4mk_BT$ and the vortex fugacity $y=e^{-E_c/k_BT}$ obey the renormalization group (RG) equations [\onlinecite{Kosterlitz74, Jose77}]
\begin{eqnarray}
\frac{d}{dl}K^{-1}(l)&=&4\pi^3y^2(l),\nonumber\\
\frac{d}{dl}y(l)&=&[2-\pi K(l)]y(l).
\label{KTRG}
\end{eqnarray}
Here $l=\ln(r/\xi)$ is the RG scale, $\xi$ is the coherence length, and $E_c$ is the vortex core energy. 

\begin{figure}
\begin{centering}
\includegraphics[width=0.6\linewidth]{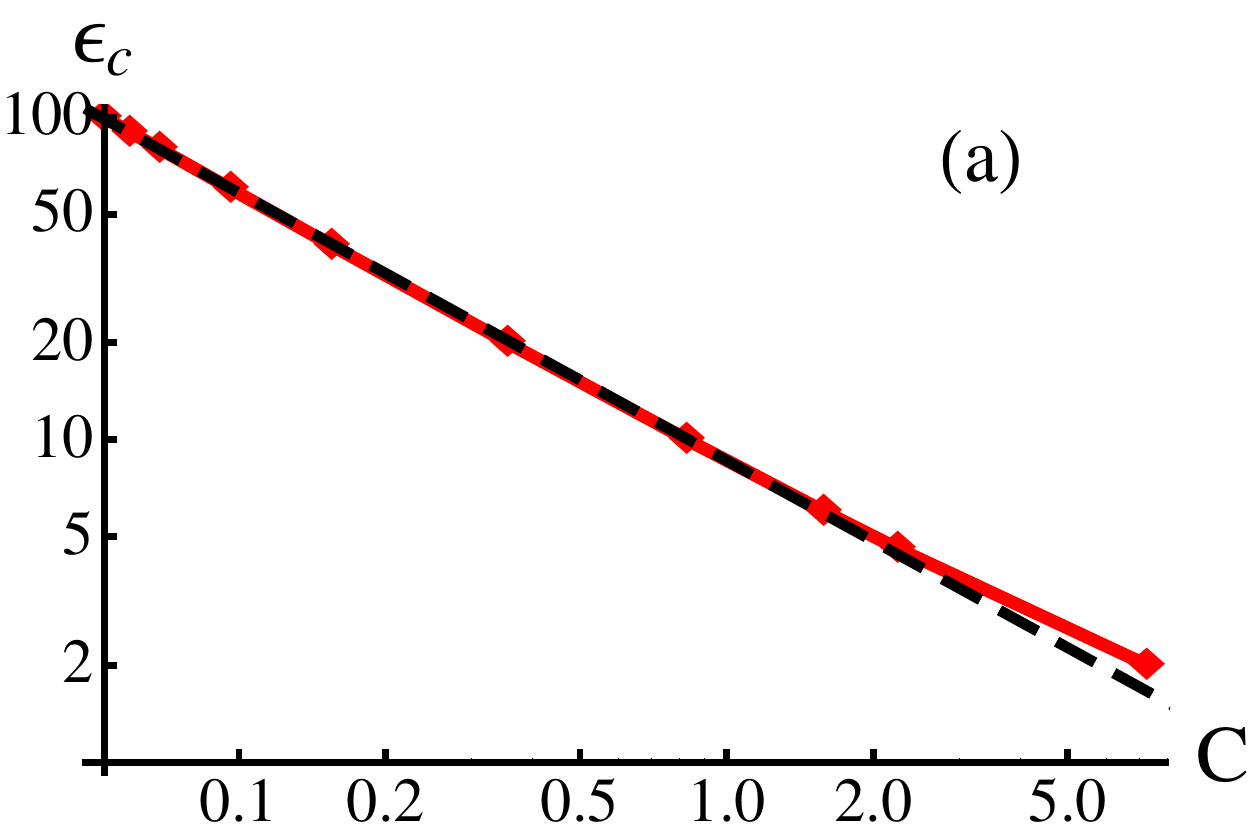}
\includegraphics[width=0.48\linewidth]{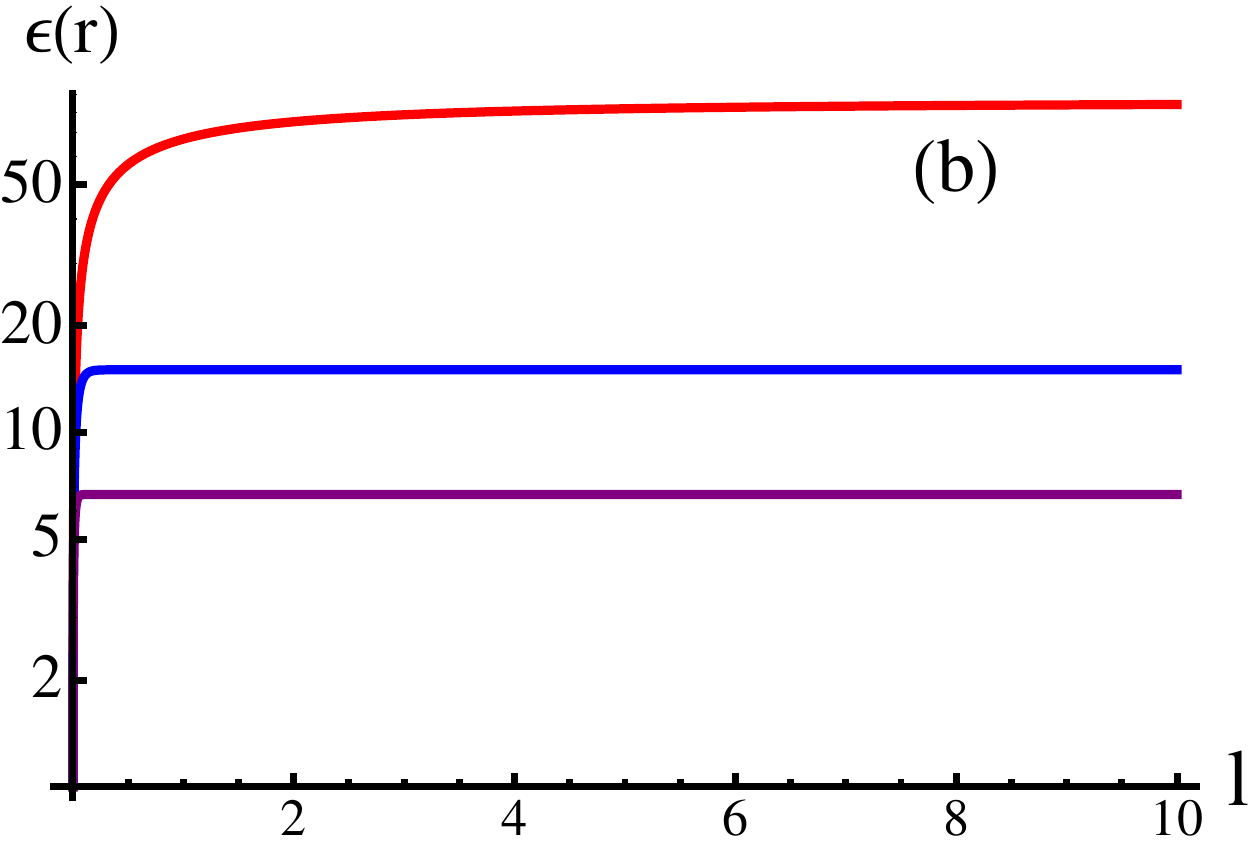}
\includegraphics[width=0.48\linewidth]{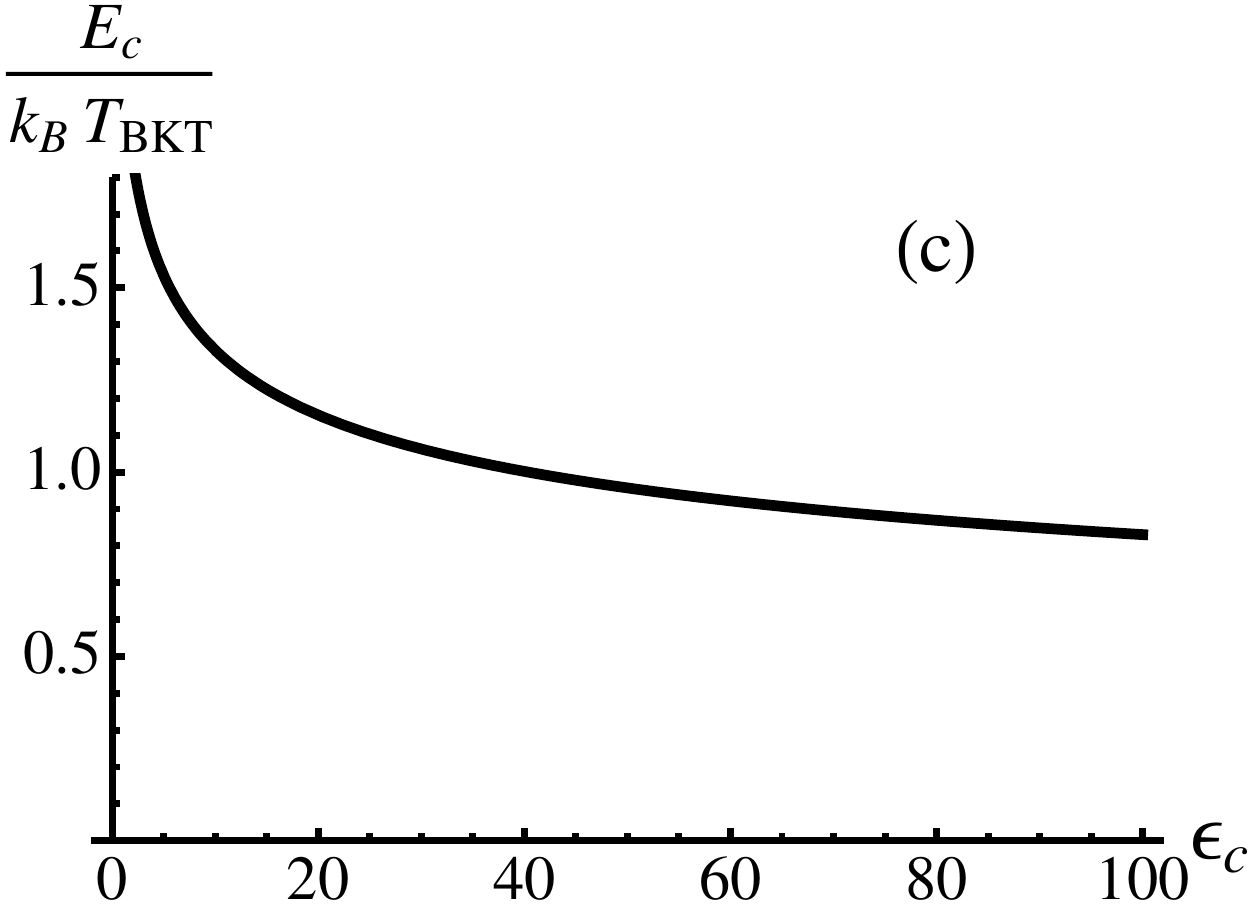}
\end{centering}
\caption{(a): The dielectric constant $\epsilon_c$ as function of the dimensionless vortex core energy $C$. The dashed line is a fit to the power law behavior. (b): Renormalization of the dielectric constant $\epsilon(r)$, for different temperatures  $T=T_{\rm BKT}, 0.95T_{\rm BKT}, 0.9T_{\rm BKT}$ (from top to bottom). Here $\epsilon_c=90, C=0.0599$. (c): The ratio of vortex core energy and BKT transition temperature as function of the dielectric constant, $E_c/k_BT_{\rm BKT}= (A^{1/\theta}/2\pi)\epsilon_c^{-(1-\theta)/\theta}$, with $\theta=0.83$. }
\label{Ec}
\end{figure}

One can define a scale-dependent dielectric constant $\epsilon(r)=K(0)/K(l)$, which measures the renormalization of the stiffness $K$ due to the screening of vortex-antivortex pairs. Without screening, $K$ takes the bulk value $K(0)=\Phi_0^2d/16\pi^3\lambda^2_{\rm b}(T)k_BT$, with $\lambda_{\rm b}$ the bulk penetration depth. Including the effect of screening, $K$ changes with the scale $r$.  One of the most important experimental consequencies of the BKT theory is that, at the BKT transition temperature, the renormalized $K$, i.e. $K(l=\infty)$, approaches a universal value  [\onlinecite{Nelson77}], which can be read out directly from the above RG equations to be $K(\infty)=2/\pi$. At $T=T_{\rm BKT}, r=\infty$, the scale-dependent dielectric constant becomes of the form $\epsilon(r=\infty,T_{\rm BKT})=\Phi_0^2d/32\pi^2\lambda^2_{\rm b}(T_{\rm BKT})k_BT_{\rm BKT}\equiv\epsilon_c$. $\epsilon_c$ is a nonuniversal number. It takes different values for different systems. For conventional superconductors, e.g. InO$_x$, it is typically 1.1 to 1.9. For YBa$_2$Cu$_3$O$_7$ thin films, it is much larger, $\epsilon_c\simeq$ 4.6 [\onlinecite{Firoy88}] or 6 [\onlinecite{Matsuda93}] . 

 The penetration depth is correspondingly renormalized with respect to the bulk value, with $\lambda^{-2}=\lambda^{-2}_{\rm b}/\epsilon(r=\infty)$. At the transition, the renormalized penetration depth satisfies the relation  [\onlinecite{Nelson77}] $k_BT_{\rm BKT} =\Phi_0^2d/32\pi^2\lambda^2$ (Eq.~(4) in the main text), which is universal in the sense that, different from $\epsilon_c$, this relation is identical for different systems. Thus to determine whether a superconducting transition is of the BKT type, it is crucial to measure the penetration depth $\lambda$, and to check whether such universal relation between $\lambda$ and $T_{\rm BKT}$ is satisfied. Such relation has been observed in superfuid helium thin films [\onlinecite{Bishop78}].

We can parameterize the vortex fugacity in term of a dimensionless quantity $C$, with $y(0)=\exp[-CK(0)/4]$ [\onlinecite{Davis90}]. $C$ is directly proportional to the vortex core energy, with $E_c=E_0C$ and $E_0=\Phi_0^2d/64\pi^3\lambda^2_{\rm b}=(\epsilon_c/2\pi)k_BT_{\rm BKT}$. The vortex core energy can be written as $E_c=(C\epsilon_c/2\pi)k_BT_{\rm BKT}$. From the above RG equations, one can see that the renormalized fugacity vanishes at the transition, i.e. $y(r=\infty, T_{\rm BKT})=0$. 

Now we proceed to quantify the relation between the vortex core energy $E_c$ (or its dimensionless counterpart $C$) and the dielectric constant $\epsilon_c$. With the initial condition $K(0)=2\epsilon_c/\pi$, $y(0)=e^{-CK(0)/4}$ and the final condition $K(\infty)=2/\pi$, $y(\infty)=0$, we can numerically solve the RG equations. We find that  $\epsilon_c=2, 4.6, 6, 90$ corresponds to $C=7.27, 2.24, 1.583, 0.0599$ respectively (see Fig.~\ref{Ec}(a)). Following the RG flow (Fig.~\ref{Ec}(b)), one can see that, only very close to the transition temperature, the dielectric constant changes substantially with scale. When moving away from $T_{\rm BKT}$, $\epsilon(r)$ quickly settles down to its infared value $\epsilon_{\infty}$, and $\epsilon_{\infty}$ decreases significantly with decreasing temperature [\onlinecite{Davis90}].

 It is interesting to notice that for $\epsilon_c\gtrsim 5$, $\epsilon_c$ and $C$ has a power law scaling, $\epsilon_c\simeq AC^{-\theta}$, with the coefficient $A\simeq 8.62$ and the power $\theta\simeq 0.83$ (see Fig.~\ref{Ec}(a)). The dielectric constant and the vortex core energy thus has the relation $\epsilon_c\simeq A(E_c/E_0)^{-\theta}$. A large dielectric constant corresponds to a small vortex core energy. For $\epsilon_c=90, C=0.0599$, the vortex core energy $E_c=(C\epsilon_c/2\pi)k_BT_{\rm BKT}\simeq (2.7/\pi) k_BT_{\rm BKT}$ \footnote{In BCS theory, the vortex core energy can be estimated as the loss of condensation energy within the vortex core, $E_c\simeq \pi \xi^2d\epsilon_{\rm cond}$, with the condensation energy density $\epsilon_{\rm cond}=N(0)\Delta^2/2$, the density of states at the Fermi level $N(0)\simeq 3n/2v_F^2m$, the BCS gap $\Delta$, and the coherence length $\xi=\hbar v_F/\pi\Delta$. Assuming $n_s=n$ at $T=0$, we have $E_c\simeq (1.9/\pi)k_BT_{\rm BKT}$ (see e.g. [\onlinecite{Mondal11}]). In XY-model, one has instead $E_c\simeq \pi k_BT_{\rm BKT}$ [\onlinecite{Nagaosa99}].
}. Taking $T_{\rm BKT}\simeq 1.6K$, one obtains $E_c\simeq 0.13 {\rm meV}$. For YBCO thin films [\onlinecite{Matsuda93}], we have $E_c\simeq (1.583\times 6/2\pi)\times 7{\rm meV}\simeq 10.6{\rm meV}$, which is one order of magnitude larger than that of heavy fermion superlattice [\onlinecite{Mizukami11}]. 

For large $\epsilon_c$, we have $E_c/k_BT_{\rm BKT}\simeq (A^{1/\theta}/2\pi)\epsilon_c^{-(1-\theta)/\theta}$  (see Fig.~\ref{Ec}(c)). Due to the small power $(1-\theta)/\theta\simeq 1/5$, for a given $T_{\rm BKT}$, a small change in the vortex core energy leads to significant change in the dielectric constant. Increasing $\epsilon_c$ from 5 to 90, the vortex core energy only changes from $1.54 k_BT_{\rm BKT}$ to $0.85 k_BT_{\rm BKT}$.

In the presence of competing orders, the vortex core energy is reduced, $E_c=E_c^{(0)}-|\delta E_c|$. As shown in the main text, $|\delta E_c|$ increases as one approaches the QCP. The dielectric constant becomes a function of the distance to the QCP, 
\begin{equation}
\epsilon_c=A\left[ \frac{E_c^{(0)}-V_0e^{-2\sqrt{a}}(3+6\sqrt{a}+4a)}{E_0} \right]^{-\theta},
\end{equation}
where $a=\alpha\lambda^4/g^2\mu_B^2\Phi_0^2$ and $\alpha$ is the distance to the QCP. $V_0$ and $a$ depends on the material specific parameters $g, \gamma$. In order to determine quantitatively the evolution of the dielectric constant near the QCP, more material specific microscopic calculations are needed.

\subsection{Effect of the interface}

 At the interface, the Yb ions disorder (due to cross diffusion and displacements) and act as nonmagnetic impurities to locally suppress superconductivity in CeCoIn$_5$ layers [\onlinecite{Bauer11}]. The superconducting order parameter is strongly suppressed near the impurity sites, and it recovers the bulk value over the distance on the order of the coherence length [\onlinecite{Franz97,Xiang95,Franz96}], $\xi(T)\simeq \nu \xi_0/\sqrt{1-T/T_{c0}}$,
with $T_{c0}$ the bulk superconducting transition temperature, $\xi_0$ the BCS coherence length, and $\nu$ a number of order unity. When the thickness of the CeCoIn$_5$ layers is large, $d>\xi(T)$, the areas of defect-depressed order parameter do not overlap, and the gap is not affected by the defects. When the thickness of CeCoIn$_5$ layers become smaller than $\xi(T)$, the depressed areas will start to overlap, and the superconducting gap in the CeCoIn$_5$ layers will be suppressed.

At low temperatures with $T\ll T_{c0}$, $\xi(T)$ is of order $\xi_0$, which is about the thickness of four layers of CeCoIn$_5$. So we expect that for $n\gg 4$, gap has the same value as the bulk material; while for $n\lesssim 4$, gap gets suppressed. This explains the experimental observation that the Pauli-limited upper critical field, which is a direct measure of the gap, retains the bulk value for $n=5,7$, and is suppressed for $n=3$.

\begin{figure}[h]
\begin{centering}
\includegraphics[width=0.48\linewidth]{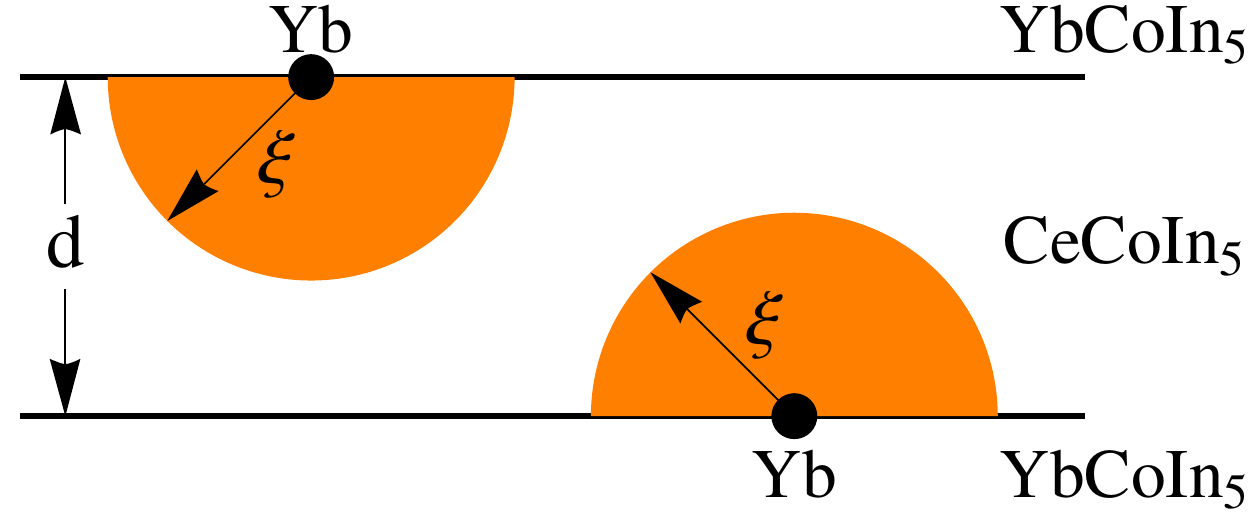}
\includegraphics[width=0.48\linewidth]{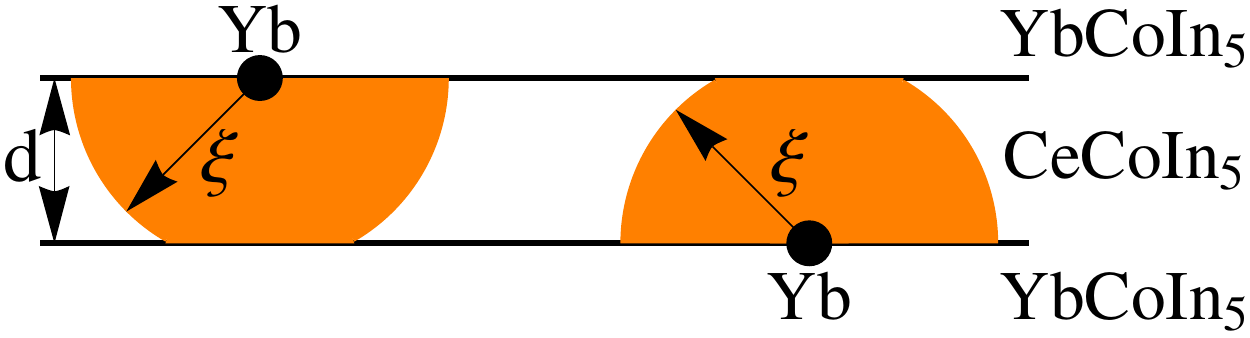}
\end{centering}
\caption{Illustration of the effect of Yb ions as pair breaking nonmagnetic impurities for $d>\xi$ and $d<\xi$. In the shaded regions of size the coherence length $\xi$ around the Yb ions, superconductivity is suppressed. }
\end{figure}

\bibliographystyle{apsrev}
\bibliography{strings,refs}

\end{document}